\documentstyle [11pt,paspconf]{article}
\input{epsf.tex}
\input{epsf.tex}
\begin{document}
\par\noindent
a contributed talk at:{{\it The International Conference on Stellar Dynamics: 
From}}
\par\hskip .2in
{\it From Classical to Modern}, Sobolev Astronomical Institute, August 2000
\vskip .2in
\sloppy%        <--- Nii
\title{Dynamics of Cuspy Triaxial Galaxies with a Supermassive Black Hole}
\author{Christos Siopis, Ioannis V. Sideris, Ilya V. Pogorelov, and
Henry E. Kandrup}
\affil{
Department of Astronomy, Department of Physics,
and Institute for Fundamental Theory,
University of Florida, Gainesville, FL 32611}
\begin{abstract}
This talk provides a progress report on an extended collaboration which has 
aimed to address two basic questions, namely: Should one expect to see cuspy, 
triaxial galaxies in nature? And can one construct realistic cuspy, triaxial
equilibrium models that are robust? Three technical results are described:
(1) Unperturbed chaotic orbits in cuspy triaxial potentials can be 
extraordinarily sticky, much more so than orbits in many other 
three-dimensional potentials. (2) Even very weak perturbations can be important
by drastically reducing, albeit not completely eliminating, this stickiness. 
(3) A simple toy model facilitates a simple understanding of why black holes 
and cusps can serve as an effective source of chaos.
%These results suggest that galaxies
%might tend initially to evolve towards configurations which, albeit not true 
%equilibria, could, in the absence of any perturbations, persist as nearly 
%stable objects over time scales long compared with the age of the Universe, 
%but which could be effectively destabilised by the effects of realistic, low 
%amplitude perturbations. 
These results suggest that, when constructing 
models of galaxies using Schwarzschild's method or any analogue thereof, 
astronomers would be well advised to use orbital building blocks that have 
been perturbed by `noise' or other weak irregularities, since such building 
blocks are likely to be more nearly time-independent than orbits evolved in 
the absence of all perturbations.
\end{abstract}
\section{Unperturbed chaotic orbits tend to be extremely sticky}
The results reported here derive from a numerical analysis of orbits
in the triaxial Dehnen potentials, where ({\it cf.} Merritt \& Fridman 
1996)
\begin{equation}
{\rho}(m)={(3-{\gamma})\over 4{\pi}abc}m^{\-{\gamma}}(1+m)^{-(4-{\gamma})},
\end{equation}
with $m^{2}=x^{2}/a^{2}+y^{2}/b^{2}+z^{2}/c^{2},$
assuming fixed axis ratios $c/a=1/2$ and $(a^{2}-b^{2})/(a^{2}-c^{2})=1/2$
but allowing for a variable cusp index $0{\;}{\le}{\;}{\gamma}{\;}{\le}{\;}2$
and a variable black hole mass 
$0{\;}{\le}{\;}M_{BH}/M_{gal}{\;}{\le}{\;}10^{-2}$.

Different segments of the same chaotic orbit can be extremely different in
terms of their visual appearance and their degree of exponential 
sensitivity (Siopis \& Kandrup 2000). These differences
can be quantified in terms of the sizes of short time Lyapunov exponents 
({\it cf.} Kandrup \& Mahon 1994) or the {\it complexity} of their
Fourier spectra, {\it i.e.,} the degree to which the power in an orbit is
concentrated near a few special frequencies ({\it cf.} Kandrup,
Eckstein, \& Bradley 1997, Siopis, Eckstein, \& Kandrup 1998). Chaotic orbits 
typically have continuous spectra,
but when they look `nearly regular' most of the power is
concentrated near a few special frequencies. In agreement with 
intuition, there is a strong correlation between the complexity of an orbit
segment and the value of its largest short time Lyapunov exponent: chaotic
orbit segments which look `nearly regular' and have less complex spectra tend 
also to exhibit comparatively small exponential sensitivity.

If two chaotic orbits in the same connected phase space region are integrated
for a sufficiently long time, it appears that they will eventually share the
same statistical properties. However, the time required for this can be
extremely long. If, for example, a single chaotic initial condition is 
integrated into the future, it can take as long as $100,000$ dynamical times
$t_{D}$, or even longer, before the short time Lyapunov exponent exhibits a
reasonable convergence towards the true Lyapunov exponent ${\chi}$, as defined
in a $t\to\infty$ limit. The overall rate of convergence can be quantified
through an examination of distributions of short time Lyapunov exponents,
$N[{\chi}({\Delta}t)]$, generated for ensembles of chaotic orbit segments of
varying length ${\Delta}t$. In the absence of any significant stickiness,
the dispersion associated with $N[{\chi}]$ scales as 
${\sigma}_{\chi}{\;}{\propto}{\;}({\Delta}t)^{-p}$ with
$p{\;}{\approx}{\;}1/2$. For very sticky orbits, $p{\;}{\ll}{\;}1/2$.

Because chaotic orbits are so sticky, one might anticipate that they could be
used as building blocks for the construction of self-consistent near-equilibria
which, albeit not strictly time-independent, behave as nearly 
time-independent entities over time intervals long compared with $t_{H}$, the
age of the Universe. (In the language of Merritt \& Fridman [1996], these
would be `quasi-equilibria' involving stochastic building blocks that are
only `partially mixed.')  However, this supposition relies crucially on the 
assumption that the statistical properties of chaotic orbit segments are
relatively insensitive to the effects of weak perturbations of the form which
act on real galaxies. In point of fact, this does not appear to be the case.
\section[]{Chaotic orbits can be surprisingly susceptible to very weak 
perturbations}
Orbits in the unperturbed triaxial Dehnen potential were perturbed to mimic 
various effects to which real stars in real galaxies are typically exposed. 
Discreteness effects, {\it i.e.,} gravitational Rutherford scattering between 
individual stars, were modeled as dynamical friction and white noise, 
{\it i.e.,} near-instantaneous kicks. The effects of one or two companion 
objects or satellite galaxies were modeled as nearly periodic perturbations. 
The effects of a dense cluster environment were modeled as coloured noise, 
{\it i.e.,} random kicks of finite duration. Internal oscillations of the form 
that might, {\it e.g.,} be triggered by a close encounter were treated as a 
superposition of normal, or pseuo-normal, modes that induced a periodic 
driving and an incoherent combination of more irregular excitations modeled as 
coloured noise.

The basic conclusion of this investigation (Siopis \& Kandrup 2000, Kandrup \& 
Siopis 2000), consistent also with analyses of motions in other two- and 
three-dimensional potentials (Pogorelov \& Kandrup 1999, Kandrup, Pogorelov, 
\& Siopis 2000), is that low amplitude irregularities can have a surprisingly 
large effect both
\par\noindent ${\bullet}$
by accelerating diffusion within a given nearly disjoint phase space
region; and 
\par\noindent ${\bullet}$ by accelerating diffusion along an Arnold web or 
through cantori connecting nearly disjoint chaotic phase space regions.
\par\noindent
Some of the topological obstructions associated with the Arnold web are 
extremely robust, so that weak perturbations have a comparatively minimal 
effect. However, in general such perturbations tend to accelerate dramatically 
the rate of phase space transport throughout the entire chaotic phase space.

The perturbations act via a resonant coupling between the characteristic
frequencies of the perturbations and the frequencies of the
orbits. That periodic driving works in this way should be 
obvious. That noise also involves a resonant coupling can be understood if one 
recalls ({\it cf.} van Kampen 1981) 
that a superposition of periodic forces 
combined with random phases is equivalent mathematically to (in general 
coloured) noise with a nonzero autocorrelation time $t_{c}$. 

Within a nearly disjoint phase space region, the perturbations allow 
microscopic motions which, in a strictly time-independent potential, are 
prohibited by Liouville's Theorem. Thus, {\it e.g.,} it becomes possible for 
phase space trajectories to cross, which helps a collection of orbits to 
`fuzz out' on short scales. The perturbations facilitate diffusion through 
cantori or along the Arnold web by `jiggling' orbits in such a fashion as to 
help them find phase space holes.

The details of the perturbation appear largely immaterial: all that seems to 
matter is the amplitude of the perturbations and their characteristic time 
scales. Even the dependence on amplitude and time scale is comparatively
weak. This implies that the details associated with realistic perturbations 
which might be difficult to extract from
observations are largely irrelevant. The overall efficacy of the perturbations 
scales logarithmically in the amplitude. For time scales $t_{c}{\;}{\gg}{\;}
t_{D}$ the perturbations have almost no effect (adiabatic limit). For somewhat
shorter time scales, the dependence on $t_{c}$ is again logarithmic.

But what amplitude is required to have a significant effect, {\it e.g.,} by
destabilising nearly time-independent building blocks? Very weak white noise
corresponding to relaxation times $t_{R}{\;}{\sim}{\;}10^{6}-10^{7}t_{D}$ can 
have appreciable effects within a time as short as $100t_{D}$, a period which,
in the inner regions of a cuspy triaxial galaxy would be short compared
with $t_{H}$. Alternatively, coloured noise and/or periodic driving 
corresponding to perturbations of fractional amplitude as small as $10^{-3}$
and a characteristic time scale $t_{c}$ as long as $10t_{D}$ can prove 
important on a time scale ${\sim}{\;}100t_{D}$. Making $t_{c}$ shorter 
facilitates a stronger resonant coupling between the perturbation and the
orbits, thus making the perturbation even more effective.

These results suggest the possibility that galaxies could settle down towards 
quasi-stationary states which, albeit not true collisionless equilibria, could 
exist as nearly time-independent entities for times ${\gg}{\;}t_{H}$, at
least in the absence of irregularities. This seems especially 
likely, given the recognition that, for triaxial systems, true equilibria will 
in general be substantially more complex than the equilibria associated with
spherical and axisymmetric configurations. For a generic triaxial system, there
is only one global integral, namely the energy $E$ or Jacobi integral $E_{J}$,
but it well known that equilibria $f(E)$ and $f(E_{J})$ cannot be used to model
triaxial systems with a strong central condensation. Unless the system is 
assumed to be characterised by a very special potential, {\it e.g.,} an
integrable Staeckel potential, it cannot be in a true equilibrium 
unless that equilibrium involves an intricate balance of `local integrals' 
(Kandrup 1998). The obvious point, then, is that even if such an intricate 
balance is hard to achieve, the system could evolve towards an approximate 
balance involving nearly time-independent building blocks. 

More pragmatically, these results would also suggest that, when constructing
equilibria using Schwarzschild's method or any analogue thereof, it would be
strongly advisable to work with an orbit library constructed from orbits that
have been evolved in the presence of weak noise or some other low amplitude
perturbations. Orbits evolved in the presence of such perturbations are more
likely to constitute nearly time-independent building blocks and, as such,
would seem less likely to be destabilised by weak irregularities associated
with discreteness effects and/or a perturbing external environment.

\section[]{Why do black holes and cusps trigger chaos?}
Numerical computations demonstrate that much of the behaviour associated 
with chaotic orbit ensembles evolved in the triaxial Dehnen potentials --
especially those associated with orbits which, in the absence of a cusp and
black hole, would correspond to regular box orbits -- can be reproduced by
the very simple potential
\begin{equation}
V(x,y,z)=
{1\over 2}(a^{2}x^{2}+b^{2}y^{2}+c^{2}z^{2}){\;}
-{GM_{BH}\over \sqrt{r^{2}+{\epsilon}^{2}}}{\;}{\equiv}{\;}V_{gal}+V_{BH},
\end{equation}
given as the sum of an anisotropic oscillator and a Plummer potential.
Orbits in this potential exhibit the same remarkable stickiness, yield 
comparable distributions of short time Lyapunov exponents, and again manifest
a strong susceptibility towards even very weak perturbations (Kandrup \&
Sideris 2000).

That this simple toy model can reproduce the qualitative features of the
more complicated potential (1) suggests strongly that {\it the results
derived for chaotic orbits in the triaxial Dehnen potentials are generic for
cuspy triaxial potentials}.
That the potential is so simple makes it comparatively easy to
understand what exactly is going on. As noted, {\it e.g.,} by Merritt (1998),
supermassive black holes in real galaxies are seldom if ever larger than
$1\%$ the mass of the entire galaxy and the central cuspy region typically
corresponds to only a small fraction of the total mass. It follows that, for
$a$, $b$, and $c$ of order unity, the physically relevant choices of $V_{BH}$
entail $M_{BH}{\;}{\ll}{\;}1$. However, the qualitative behaviour in this
regime is easily understood by a combination of perturbation theory and
common sense.

For $M_{BH}{\;}{\ll}{\;}1$ and ${\epsilon}\to 0$ in eq. (2), it appears that, 
except for the very lowest energies (where the potential is essentially 
Keplerian), essentially all of the orbits are chaotic, but that 
they tend to behave in a nearly regular fashion nearly all of the time. As
the orbit evolves it will 
usually find itself in a region where $|V_{BH}|{\;}{\ll}{\;}|V_{gal}|$, 
so that the potential in which it is moving is very nearly integrable and the 
short time Lyapunov exponents are extremely
small. Occasionally, however, the orbit will move comparatively close to the
center of the galaxy, so close that $|V_{BH}|$ becomes comparable to 
$|V_{gal}|$. When this happens the orbit feels the competing influences of 
two different potentials of comparable magnitude with very different 
symmetries and the values of
the positive short time Lyapunov exponents increase precipitously. The fact
that, for small $M_{BH}$, almost all the orbits are chaotic is not difficult
to understand. In the limit that
 $M_{BH}{\;}{\equiv}{\;}0$, the orbits all reduce to boxes
which densely fill a region in configuration space that includes the origin.
One might expect that, for small but nonzero $M_{BH}$, the orbits can still
pass arbitrarily close to the origin but, for any nonzero $M_{BH}$ there is
a minimum radius $r_{min}$ inside of which $|V_{BH}|$ becomes large compared
with $|V_{gal}|$.

\begin{figure}[t]
\centering
\centerline{
        \epsfxsize=9cm
        \epsffile{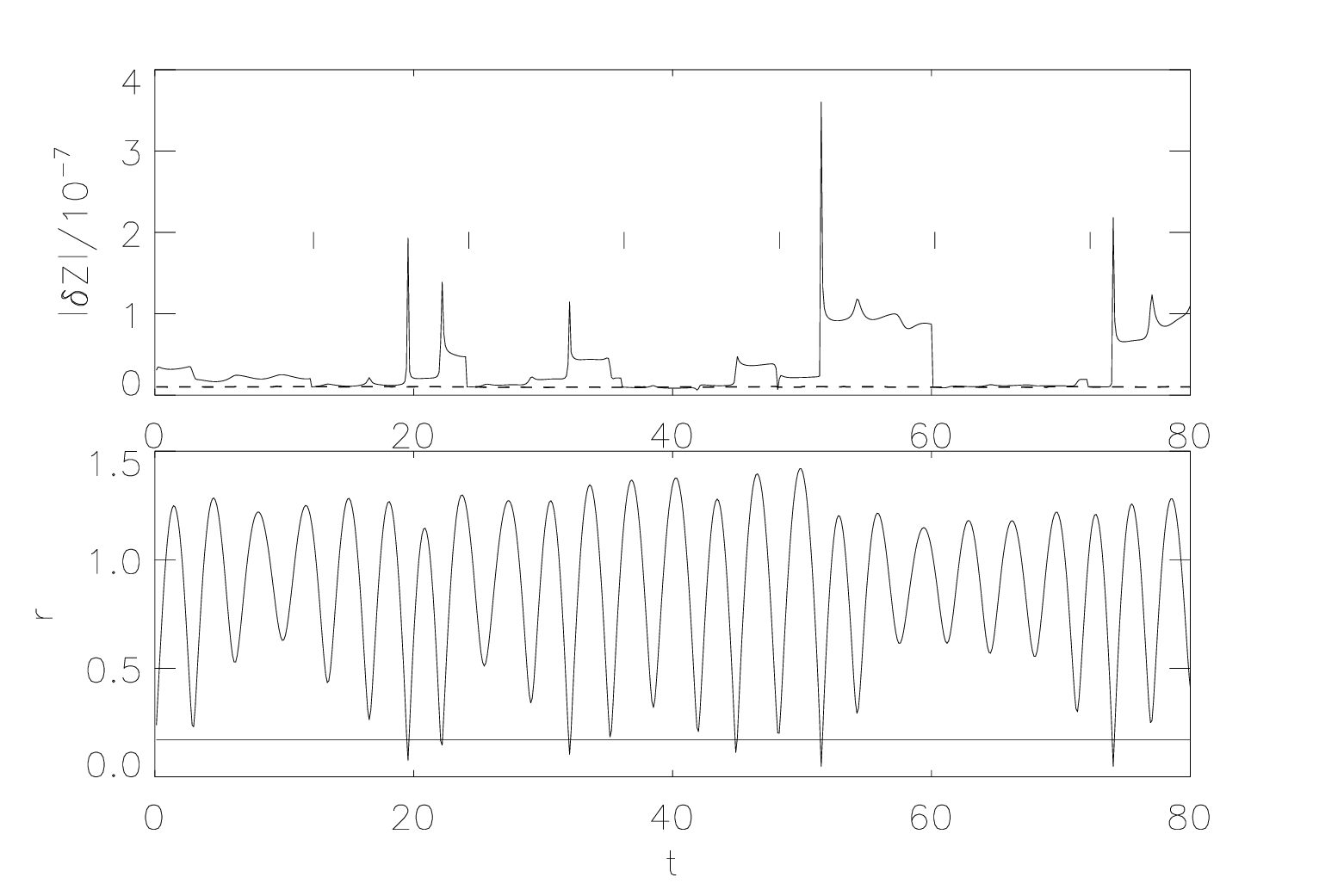}
           }
        \begin{minipage}{12cm}
        \end{minipage}
        \vskip -0.4in\hskip -0.0in
\caption{ %FIG. 1 
\small
(a)
The solid curve exhibits $|{\delta}Z(t)|$, the phase space distance between
a perturbed and unperturbed chaotic orbit with $E=0.75$ evolved in the 
potential (2) for $a^{2}=1.25$, $b^{2}=1.0$, $c^{2}=0.75$, $M_{BH}=0.15$, and 
${\epsilon}^{2}=10^{-4}$. 
The dashed line shows $|{\delta}Z|$ for a regular orbit evolved in
the same potential with the same energy. The perturbation was renormalised
at intervals ${\delta}t=12$ at the points indicated by vertical stripes. (b)
$r(t)$, the distance from the origin for the same unperturbed orbit at the
sames times. The vertical line corresponds to $r=0.17$, for which $|V_{BH}|
{\;}{\approx}{\;}0.88$.} \label{fig-1}
\vspace{-0.2cm}
\end{figure}

That significant chaos is only triggered when the trajectory passes relatively
close to the black hole is illustrated in Figure~\ref{fig-1}, which
%\begin{figure}
%\vspace{6cm}
%\caption{(a)
%The solid curve exhibits $|{\delta}Z(t)|$, the phase space distance between
%a perturbed and unperturbed chaotic orbit with $E=0.75$ evolved in the 
%potential (2) for $a^{2}=1.25$, $b^{2}=1.0$, $c^{2}=0.75$, $M_{BH}=0.15$, and 
%${\epsilon}^{2}=10^{-4}$. 
%The dashed line shows $|{\delta}Z|$ for a regular orbit evolved in
%the same potential with the same energy. The perturbation was renormalised
%at intervals ${\delta}t=12$ at the points indicated by vertical stripes. (b)
%$r(t)$, the distance from the origin for the same unperturbed orbit at the
%sames times. The vertical line corresponds to $r=0.17$, for which $|V_{BH}|
%{\;}{\approx}{\;}0.88$.} \label{fig-1}
%\end{figure}
exhibits a segment of a chaotic orbit % with $E=0.75$ 
evolved in the potential (2).
% with $a^{2}=1.25$, $b^{2}=1.0$, $c^{2}=0.75$ and $M_{BH}=0.15$. The solid 
curve in the top panel exhibits the phase space separation $|{\delta}Z|$ 
between the original orbit and a perturbed orbit displaced originally by a 
distance $|{\delta}Z|= 10^{-8}$ and periodically renormalised in the usual way 
({\it cf.} Lichtenberg \& Lieberman 1992). % at intervals ${\delta}t=12$. 
The 
dashed curve shows an analogous plot of $|{\delta}Z|$ for a regular orbit. The 
lower panel plots $r(t)$, the distance from the origin. Most of the time the
perturbed and unperturbed orbits remain very close together, with comparatively
little systematic exponential divergence. Only when $r$ becomes as small as 
${\sim}{\;}0.17$, so that $|V_{BH}|$ becomes as large as ${\sim}{\;}0.88$,
do the orbits tend to diverge significantly. 

This toy model is particularly simple since, in the limit $M_{BH}\to 0$, all
the orbits are regular boxes. For 
generic triaxial potentials, in the absence of a cusp or black hole one
would expect both centrophilic box orbits and centrophobic tubes. Because
the tubes are centrophobic, they should not in general be impacted all that
much by the introduction of a cusp or a central black hole. What, {\it does}, 
however, seem to be true is that many of the orbits which, in the absence of
a cusp, behave as regular boxes can, in the presence of a cusp or black hole,
be converted into orbits which, albeit formally chaotic, behave in a nearly
regular fashion much of the time.
\acknowledgments
We are pleased to acknowledge useful collaborations with Brendan Bradley,
Barbara Eckstein, Salman Habib, and Elaine Mahon.
This research was supported in part by NSF AST-0070809 and by the Institute
for Geophysics and Planetary Physics at Los Alamos National Laboratory.

%\vfill\eject

\begin{references}
\reference Kandrup, H. E. 1998, \mnras, 299, 1139
\reference Kandrup, H. E., Eckstein, B. L., Bradley, B. O. 1997, A\&A, 320, 65
\reference Kandrup, H. E., Mahon, M. E. 1994, A\&A, 290, 762
\reference Kandrup, H. E., Pogorelov, I. V., Sideris, I. V. 2000, \mnras, 
311, 719
\reference Kandrup, H. E., Sideris, I. V. 2000, \mnras, submitted
\reference Kandrup, H. E., Siopis, C. 2000, in preparation
\reference Lichtenberg, A. J., Lieberman, M. A. 1992, Regular and Chaotic
Dynamics. Springer: Berlin
\reference Merritt, D. 1998, Comments Astrophys., 19, 1
\reference Merritt, D., Fridman, T. 1996, \apj, 460, 136
\reference Pogorelov, I. V., Kandrup, H. E. 1999, Phys. Rev. E, 60, 1567
\reference Siopis, C., Kandrup, H. E. 2000, \mnras, in press
\reference Siopis, C., Eckstein, B. L., Kandrup, H. E. 1998, Ann. N. Y. Acad.
Sci., 867, 41
\reference van Kampen, N. G. 1981, Stochastic Processes in Physics and
Chemistry. North Holland: Amsterdam
\end{references}
\end{document}